\documentclass[5p,twocolumn]{elsarticle}

\usepackage{braket}
\usepackage{amsmath}
\usepackage{booktabs} 

\newcommand{\spinup}{\uparrow}
\newcommand{\spindown}{\downarrow}
\newcommand{\lll}{{(l)}}
\newcommand{\llp}{{(l')}}
\newcommand{\mll}{{(-l)}}

\begin{document}

\title{Computation of ground-state properties of strongly correlated
  many-body systems by a two-subsystem ground-state approximation}

\author[tugrazitp]{Ralf Gamillscheg}
\ead{ralf.gamillscheg@tugraz.at}

\author[kfmath]{Gundolf Haase}

\author[tugrazitp]{Wolfgang von der Linden}

\address[tugrazitp]{Institute of Theoretical Physics - Computational
  Physics, Graz University of Technology, Graz, Austria}
\address[kfmath]{Institute for Mathematics and Scientific Computing, Karl-Franzens-University Graz, Graz, Austria}

\date{\today}

\begin{abstract}
  We present a new approach to compute low lying eigenvalues and
  corresponding eigenvectors for strongly correlated many-body
  systems. The method was inspired by the so-called Automated
  Multilevel Sub-structuring Method (AMLS). Originally, it relies on
  subdividing the physical space into several regions. In these
  sub-systems the eigenproblem is solved, and the regions are combined
  in an adequate way. We developed a method to partition the state
  space of a many-particle system in order to apply similar operations
  on the partitions. The tensorial structure of the Hamiltonian of
  many-body systems make them even more suitable for this approach.
  
  The method allows to break down the complexity of large many-body
  systems to the complexity of two spatial sub-systems having half the
  geometric size. Considering the exponential size of the Hilbert
  space with respect to the geometric size this represents a huge
  advantage. In this work, we present some benchmark computations for
  the method applied to the one-band Hubbard model.
\end{abstract}


\maketitle

\section{Introduction}
In modern solid-state physics strongly correlated many-body systems
play an increasingly important role, as e.g. in the case of the high
temperature superconductors, the manganites and vanadates, and more
recently also in light-matter- and ion trap quantum simulators.  The
numerical problem of addressing strong correlations leads to
eigenvalue problems of matrices whose size depend exponentially on the
geometric size and the particle number.

Several methods have been introduced to solve problems of this
kind. \emph{Quantum Monte-Carlo} (QMC, \cite{VonDerLinden_QMC}) is a
powerful method to deal with finite-temperature systems, where large
system sizes can be reached. On the downside, many system
configurations cannot be addressed efficiently due to the so-called
\emph{sign-problem} \cite{evertz_LOOP}. Another powerful method is the
\emph{Density Matrix Renormalization Group} (DMRG,
\cite{Schollwoeck_DMRG}) which allows to solve for ground-state
properties of comparatively large-scale electronic structures. Its
disadvantages include the lack of possibility of treating problems
apart from 1D or pseudo-1D problems. Newer methods like the
\emph{Variational Cluster Perturbation Theory (VCPT)} also deliver
promising results.

For this work we followed a new approach to break down the numerical
complexity of such systems. Rather than starting from a physical point
of view we let ourselves be inspired by other fields of numerical
simulations, which also deal with large-scale eigenvalue
calculations. The \emph{Automated Multilevel Sub-structuring Method} (AMLS,
\cite{bekas_amls,bennighof_amls_2004}) e.g. for linear elastodynamics
relies on partitioning the physical space into smaller pieces wherein
the eigenproblem is solved as a starting point for the calculation of
the full system eigenvalues. We refer to our approach as Two
Sub-system Ground-state Approximation (TSGSA).

\section{Partitioning the occupation number state space}

In second quantization the ab initio and approximation free form of the Hamiltonian for the electronic degrees
of freedom reads
\begin{align*}
  \hat H &= \sum_{\nu,\nu'} t_{\nu \nu'} \hat a^\dagger_\nu \hat
  a^{\phantom{\dagger}}_{\nu'} + \sum_{\nu,\nu',\mu,\mu'} V_{\nu \nu'
    \mu \mu'} \hat a^\dagger_\nu
  \hat a^\dagger_{\mu}\hat a^{\phantom{\dagger}}_{\nu'}
  \hat a^{\phantom{\dagger}}_{\mu'} \;,
\end{align*}
where the operator $\hat a^\dagger_\nu$ ($\hat a^{\phantom{\dagger}}_\nu$)
creates (annihilates) a fermion in orbital $\phi_\nu$. The quantity
$\nu=(i,\alpha,\sigma)$ represents a combined index including the
site- (unit cell-) index $i$, the index $\alpha$ of the basis function
to describe the orbitals of the various atoms within a unit cell, and
the spin $\sigma$. $t$ and $V$ are the corresponding matrix elements
for the single particle part (hopping) and the interaction,
respectively. A suitable basis for the many-body problem in second
quantization is the occupation basis $|\{n_\nu\}\rangle$, where the
occupation $n_\nu$ for fermions can only be 0 or 1, in accordance with
the Pauli principle. There are three sources for the intricacy of a
many-body problem. The first one is the number of orbital degrees of
freedom per unit cell, required for a quantitatively accurate
description of the band structure. The second concerns the
overwhelming number of basis states for a true many-body
calculation. Let $L_r$ be the number of unit cells and $L_b$ the
number of orbitals within a unit cell then there are $L=L_r L_b $
different index tuples $\nu=(i,\alpha)$ for each electron spin
direction. There are correspondingly $L$ occupation numbers $n_\nu$
which can either be 0 or 1. The number of many-body basis states to
describe a system of $N_\uparrow$ ($N_\downarrow$) electrons with spin
up (down) is given by the number of possibilities to distribute
$N_\uparrow$ entries 1 and $N-N_\uparrow$ entries 0 among the $L$
occupation numbers and likewise for the spin down electrons. Hence the
number $M$ of many-body basis states is
\begin{align*}
    M = {L\choose N_\uparrow}{L\choose N_\downarrow}\;.
\end{align*}
It is needless to emphasize that a true many-body ab-initio
calculation is out of reach.

In order to study the generic properties of strongly correlated
many-body systems qualitatively it is, however, sufficient to reduce
the number of orbital degrees of freedom to a minimum. Common
many-body models include up to $L_b=3$ orbitals per unit cell.

But even with a strongly reduced number of orbital degrees of freedom
there remains a third problem, the structure of the interaction part,
which is still too complicated for an exact treatment, as well by
numerical as by analytical means. There is reason to believe that the
genuine many-body effects can already by described and understood when
only short ranged density-density terms of the form
\begin{align*}
    \hat H_{\text{int}} &= \sum_{\nu\nu'} V_{\nu\nu'} \hat n_{\nu} \hat n_{\nu'}\;,
\end{align*}
are retained in the model. The density operator is given by $\hat n_{\nu}=\hat a^\dagger_\nu \hat a^{\phantom{\dagger}}$.

Though not really necessary, but in order to keep the number of
parameters small, the hopping part is commonly approximated by a
tight-binding form, allowing for nearest neighbor hopping only. The
following \emph{extended Hubbard-model} with one orbital degree of
freedom per unit cell includes an on-site and a next-nearest
neighbor interaction:
\begin{align*}
     \hat H &= -t\sum_{\langle i,j\rangle,\sigma} \hat a^\dagger_{i,\sigma} \hat a_{j,\sigma}
     + U \sum_i \hat n_{i,\uparrow} \hat n_{j,\downarrow} + V \sum_{\langle i,j\rangle} \hat n_i \hat n_j\;.
 \end{align*}

 Here, we will consider the two most simple and common fermionic
 models, the case of spin-less fermions (only one spin-species) with
 nearest neighbor interaction and the Hubbard model (only on-site
 Coulomb interaction).

 For spin-less fermions the Hamiltonian reads
\begin{align*}
    \hat H &= -t \sum_{\langle i j\rangle} \hat a^\dagger_i \hat a^{\phantom{dagger}}_j + V \sum_{\langle i j\rangle} \hat n_i \hat n_j\;,
\end{align*}
where $i$ stands for the unit cells and $\langle i j\rangle$ indicates that the unit cells $i$ and $j$ are nearest neighbors.

The Hubbard Hamiltonian is given by
\begin{align*}
    \hat H &= -t \sum_{\langle i j\rangle,\sigma} \hat a^\dagger_{i\sigma} \hat a^{\phantom{dagger}}_{j\sigma} + U \sum_{i} \hat n_{i\uparrow} \hat n_{i\downarrow}\;.
\end{align*}

Despite of the strong reduction in degrees of freedom, the number of
many-body basis states is still very large and increases exponentially
with increasing number of orbitals (sites) $L$.  In the case of
spin-less fermions we have $M={L\choose N}$ and for the Hubbard model,
it reads as mentioned before $M = {L\choose N_{\uparrow}}{L\choose
  N_{\downarrow}}$.

For example, a system of spin-less fermions on $L=20$ sites with
$N=10$ electrons has $M=184,756$ many-body basis states, while a system
of half the size, i.e. $L=10$ sites with $N=5$ has merely $M=252$. For
the Hubbard model the situation is even more pronounced. Again for a
half filled system with 20 sites and 10 electrons of each species, we
find $M\approx 3 \cdot 10^{10}$, while the half filled system on 10
sites has $M\approx 6 \cdot 10^4$. So the numerical complexity increases by
roughly $10^6$ if ´we double the geometrical size of the system.

The idea of the present paper is to exploit systematically the fact that smaller sub-systems have a significantly reduced Hilbert space.

\subsection{Sectors in occupation number state space}

The full Hamiltonian, be it the Hubbard or the spin-less fermion
model, conserves the number of particles per spin direction
$N_\sigma$. Therefore, the Hamilton matrix is block diagonal in the
occupation number basis due to the conservation of particle numbers
and we solve the eigenvalue problem for fixed number of particles
$(N_\uparrow,N_\downarrow)$ separately.  The case of the spin-less
fermion model can be traced back to the Hubbard case by assuming only
one spin direction. Here we restrict the discussion to an even number
of lattice sites $L=2 L_h$ and an even number of electrons per spin
direction $N_\sigma = 2 N_{h,\sigma}$.  Although the generalization is
straightforward, it would hamper the discussion unnecessarily.

Now we split the lattice into two sub-systems $A$ and $B$ of equal
size. In the approach to be discussed below, we will start from the
case of completely decoupled sub-systems, i.e. all hopping and
interaction effects between the two sub-systems are ignored.
Consequently, the numbers of electrons per spin direction are
conserved in each sub-system separately.  Due to the particle number
conservation in the sub-systems, the corresponding occupation number
state space can be split into different sectors
$(l_\uparrow,l_\downarrow)$, which are characterized by the number of
particles per spin direction $(N_{\uparrow}=N_{h,\uparrow}
+l_\uparrow,N_\downarrow = N_{h,\downarrow} +l_\downarrow)$ in the
sub-systems under consideration.  Here we have introduced a the
quantity $l_\sigma$ which specifies the deviation of the actual
particle number for spin $\sigma$ from the reference value
$N_{h,\sigma}$. Obviously, $l_\sigma$ can range from $-N_{h,\sigma}$
to $+N_{h,\sigma}$.

Next we construct a complete basis for the entire system, by
tensor-products of the eigenvectors of the decoupled sub-systems.  In
order to achieve the correct particle numbers per spin direction for
the entire system, only specific sectors
$(l_{\alpha,\uparrow},l_{\alpha\downarrow})$ of the two sub-systems
can be combined, $(l_\uparrow,l_\downarrow)$ for sub-system $A$ and
$(-l_\uparrow,-l_\downarrow)$ for sub-system $B$. It will turn out
expedient to introduce the Manhattan distance in the
$(l_{\alpha,\uparrow},l_{\alpha\downarrow})$-plain, which is given by
$d( (l_{\alpha,\uparrow},l_{\alpha\downarrow}) = |l_{\alpha,\uparrow}|
+ |l_{\alpha,\downarrow}|$. Sectors of the same Manhattan distance
form shells, which are of equal importance as far as eigenvectors and
eigenvalues are concerned.

\begin{figure}
\centering
\includegraphics[height=6cm]{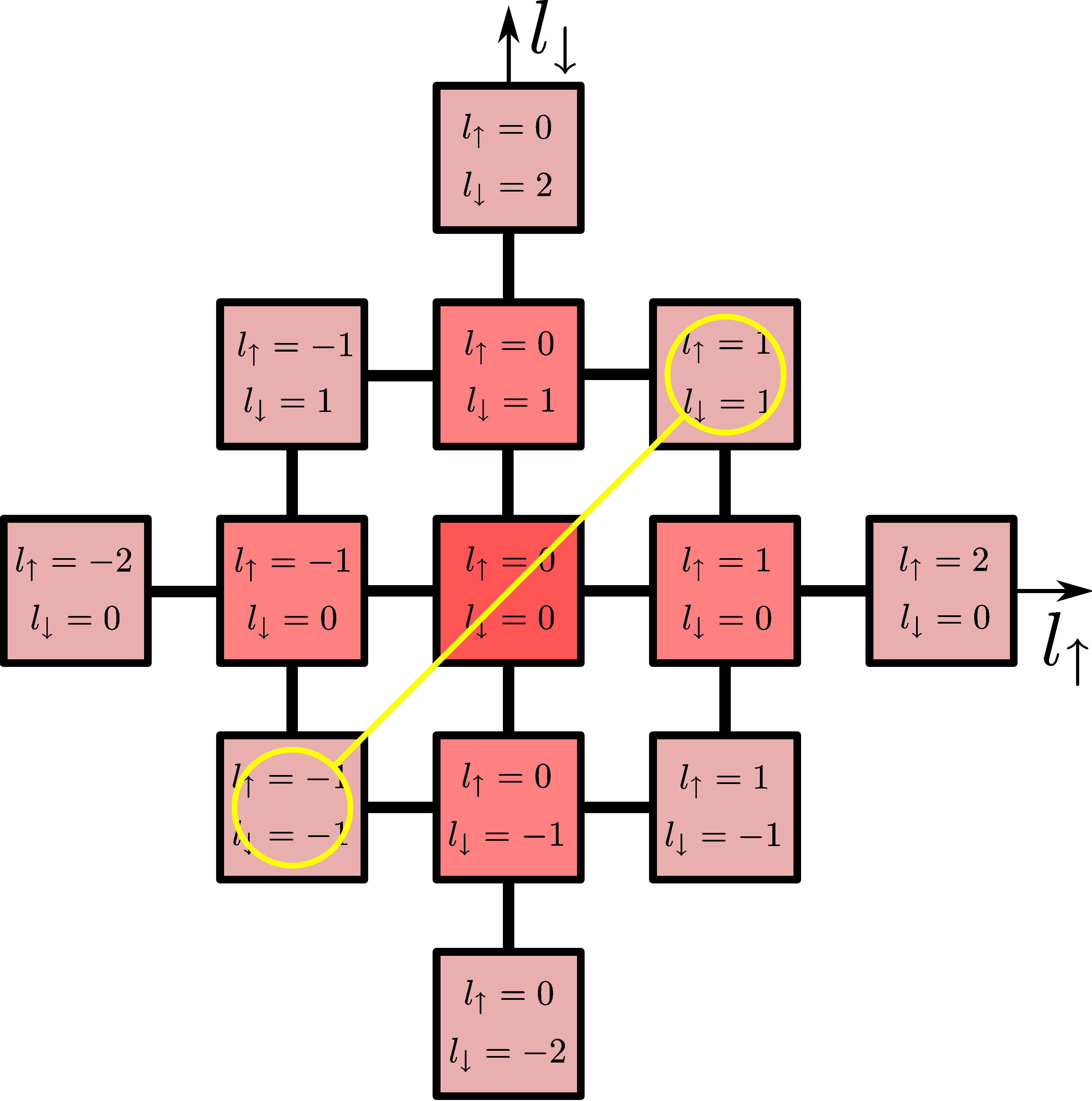} 
\caption{Particle number sectors ($(l_\uparrow,l_\downarrow)$) for a
  Hubbard system.  Shells of sectors with the same Manhattan have the same
  color.  The yellow line marks a possible combination of sectors to
  build a
  partition.\label{fig:TLGSA_subsystems_hubbard_with_combination}}
\end{figure}

In fig. \ref{fig:TLGSA_subsystems_hubbard_with_combination}
different shells up to to Manhattan distances 2 are depicted.  As
pointed out before, within one shell there are always two opposite
$(l_\uparrow,l_\downarrow)$-pairs, one for each sub-system, which form
a particle number partition which is used in the tensor-product basis
for the full system. One such example is indicated in the figure by the
yellow line.  The central square denotes the sector with Manhattan
distance 0, i.e. ($l_\spinup=l_\spindown=0$).

\subsection{Hamilton operator of the sub-systems}
It is expedient to introduce a combined site index $i\rightarrow (\alpha,i)$,
where $\alpha \in\{A,B\}$ refers to the sub-systems and and henceforth
$i\in\{1,\ldots,N_h\}$ enumerates the sites within each sub-system.
The Hamiltonian can be decomposed into three parts
\begin{align*}
    \hat H &=  \underbrace{\hat H_A + \hat H_B }_{:=\hat H_0}+ \hat H_{AB}\;.
\end{align*}
For spin-less fermions the parts of the Hamiltonian are
\begin{align*}
    \hat H_\alpha &=& -t \sum_{\braket{i,j}}
    \hat a^\dagger_{\alpha,i} \hat a^{\phantom{dagger}}_{\alpha,j} + V \sum_{\braket{i,j}} \hat n_{\alpha,i} \hat n_{\alpha,j}\;;\quad \alpha\in\{A,B\}\\
        \hat H_{AB} &=& \underbrace{-t  {\sum_{i,j}}' \bigg(\hat a^\dagger_{A,i} \hat a^{\phantom{dagger}}_{B,j} + \hat a^\dagger_{B,j} \hat a^{\phantom{dagger}}_{A,i}\bigg)}_{ =:\hat H_{AB}^\text{kin}} + \underbrace{V{\sum_{i,j}}' \hat n_{A,i} \hat n_{B,j} }_{    =:    \hat H_{AB}^\text{int}}
\end{align*}
where $\sum'$ indicates that the indices $i,j$ have to be chosen such that $(A,i)$ and $(B,j)$ belong to nearest neighbor sites.

For the Hubbard model the splitting yields
\begin{align*}
    \hat H_\alpha &= -t \sum_{\braket{i,j},\sigma}
    \hat a^\dagger_{\alpha,i,\sigma} \hat a^{\phantom{dagger}}_{\alpha,j,\sigma} + U \sum_{i} \hat n_{\alpha,i\uparrow} \hat n_{\alpha,i\downarrow}\;;\;\alpha\in\{A,B\}\\
    \hat H_{AB} &= -t  {\sum_{i,j,\sigma}}'  \bigg(  \hat a^\dagger_{A,i,\sigma} \hat a^{\phantom{dagger}}_{B,j,\sigma} +
    \hat a^\dagger_{B,j,\sigma} \hat a^{\phantom{dagger}}_{A,i,\sigma} \bigg)
\end{align*}
In case of the spin-less fermion model there is a two-particle
coupling term between the two sub-systems ($\hat H_{AB}^\text{int}$) which
is detrimental for cluster perturbation theory (CPT or VCA,
\cite{Senechal_CPT,Potthoff_VCA}). In the present approach it does not
make any difference at all.

Now, the particle numbers in the two sub-systems are not conserved, due
to the inter-sub-system hopping, and in principle all particle numbers
between $0$ and $N$ are conceivable for each sub-system. However, the
most probable number of particles in the sub-systems is $N_A = N_B =
N_h$. The particle number fluctuations in the two sub-systems is
subject to the constraint $N_A+N_B=N$ and likewise for the two spin
species $N_{A,\sigma} + N_{B,\sigma} = N_{h,\sigma}$.  If the two
sub-systems are decoupled, i.e. for $\hat H_{AB}=0$ the particle numbers in
each sub-system are conserved as well.

\begin{figure}
\centering
\includegraphics[width=7cm]{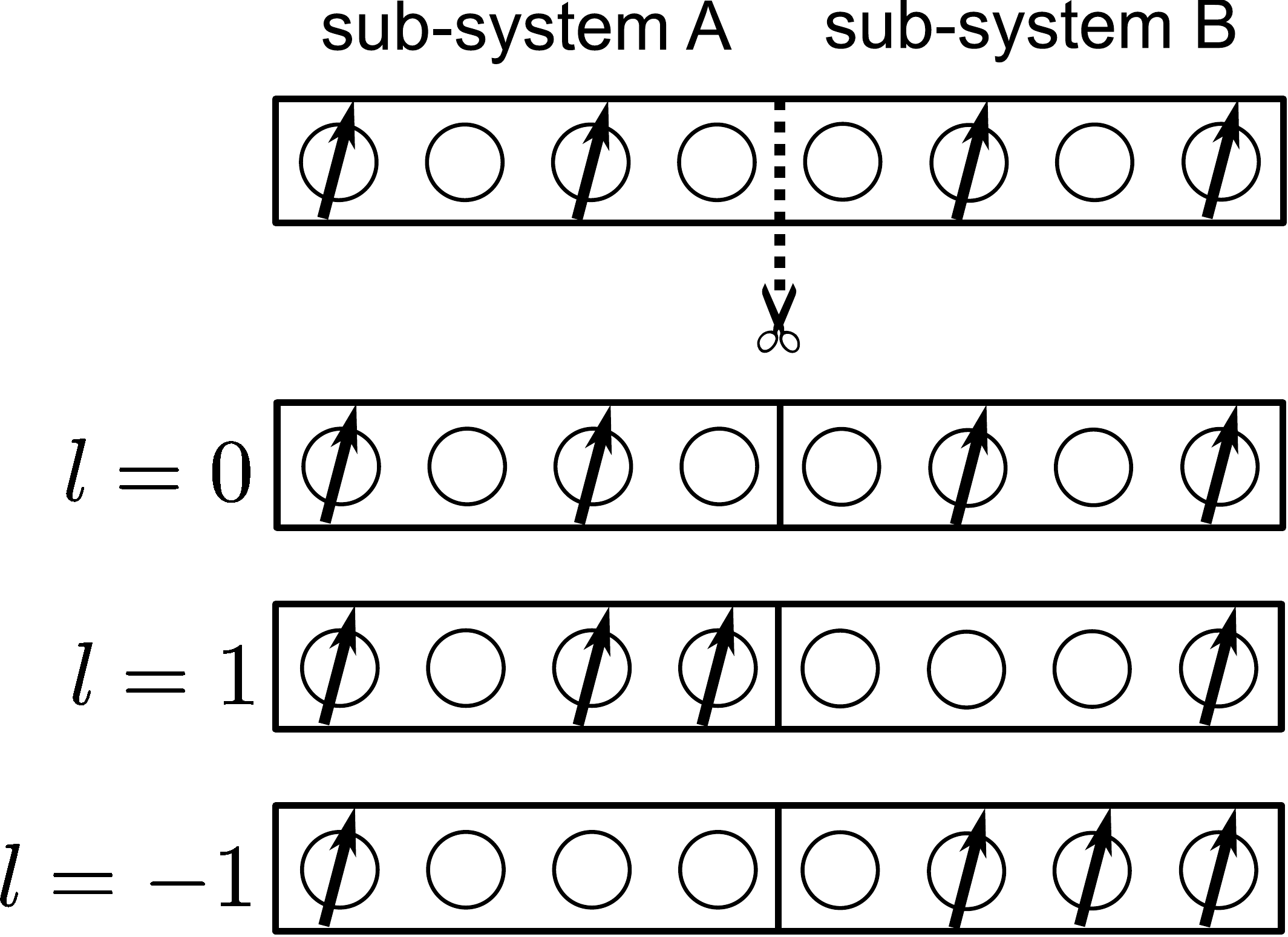}
\caption{Partitioning the occupation number state space by number of
  particles in each sub-system. Here, representatives of each partition
  are depicted.\label{fig:TLGSA}}
\end{figure}

In the case of the Hubbard model the partitioning concerns both spin
species separately. A partition $(l_\uparrow,l_\downarrow)$ therefore
represents the situation of $N_{A,\sigma} = N_{h\sigma} + l_\sigma$
and $N_{B,\sigma} = N_{h\sigma} - l_\sigma$, respectively.  Typical
configurations for partitions with $l=0$ and $l=\pm1$ are depicted in
fig. \ref{fig:TLGSA}.

The eigenvectors of $\hat H_0$ are simply tensor products of the
eigenvectors of the separate sub-system Hamiltonians $\hat H_\alpha$
and the eigenvalues are given by the sum of the corresponding
eigenvalues.  Let the eigenvalue problem of sub-system $\alpha$ with
$N_\alpha=N+l$ electrons be given by
\begin{align*}
    \hat H_\alpha \ket{\Psi_{\nu}^\lll}_\alpha &= E^\lll_{\nu} \ket{\Psi_\nu^\lll}_\alpha\,.
\end{align*}
Here the meaning of the the indices is as follows. The lower outer
index $\alpha$ stands for the sub-system, the upper index $\lll$
represents the particle number ($N_\alpha = N_h + l$), and the lower
index $\nu$ enumerates the eigenvalues and eigenvectors for this
particle number. Note, that the eigenvalue spectrum is the same for
both sub-systems if they are occupied by the same number of electrons.

The eigenvectors and eigenvalues of $H_0$ in partition $l$ are given
by
\begin{align*}
\ket{\psi^\lll_{\nu}} &:= \ket{\Psi_{\nu_A}^\lll}_A \otimes \ket{\Psi_{\nu_B}^\mll}_B\\
E^\lll_{\nu} &:= E^\lll_{\nu_A}  + E^\mll_{\nu_B} \;,
\end{align*}
where the index $\nu$ represents the entire system with $\nu =
(\nu_A,\nu_B)$.

Due to the missing coupling of the sub-systems, the eigenvalue problem
of $\hat H_0$ has a strongly reduced complexity.  If we add $\hat H_{AB}$ by
perturbation theory, there will be a first order contribution only
from the interaction term in the spin-less fermion case.  The hopping
terms do not contribute in first order, as they change the number of
particles in the sub-systems.

The ground-state of $\hat H_0$ is obtained for $N_A=N_B=N_h$, i.e. it belongs to partition $l=0$. The second order energy correction is given by
\begin{align*}
    \Delta E^{(2)} &= \sum_{l=-N_h}^{N_h} {\sum_{\nu}}'\frac{
    \big|
    \langle \psi^{(l)}_\nu |\hat H_{AB}|\psi^{(0)}_0 \rangle
    \big|^2
    }{
    E^{(l=0)}_0 -E^{(l)}_\nu
    }\;.
\end{align*}
As usual $\sum'$ indicates that $(l=0\wedge \nu=0)$ is excluded from
the sum.  Only $l=0$ and $l=\pm 1$ contribute to the energy
correction, because only one electron can hop at a time across the
border .  In other words, the partitions $l=\pm 1$ come into play. The
importance of the unperturbed eigenvectors of partitions $l=\pm 1$ is
determined by the matrix the element $M_{\nu}^{(l=\pm 1)}:=\langle \psi^{(l=\pm
  1)}_\nu |\hat H_{AB}|\psi^{(0)}_0\rangle $ and the inverse of the
energetic distance $E^{(l=0)}_0 -E^{(l)}_\nu$. The latter implies that
low eigenstates of the unperturbed system are more important for the
ground-state of the entire system than those with higher energies. In
addition the kinetic coupling matrix element is driven by the hopping
across the border and yields for the spin-less fermion model
\begin{align*}
    M_{\nu}^{(l=\pm 1)} &= -t {\sum_{i j}}'\langle \psi^{(\pm 1)}_\nu |
    \big(
    \hat a^\dagger_{A,i} \hat a^{\phantom{\dagger}}_{B,j} + \hat a^\dagger_{B,j} \hat a^{\phantom{\dagger}}_{A,i}
   \big)|\psi^{(0)}_0 \rangle\\
       M_{\nu}^{(l=+ 1)} &= -t{\sum_{ i j}}'
       \langle \psi^{(+1)}_{\nu_A} |\hat a^\dagger_{A,i} |\psi^{(0)}_{0} \rangle_A
       \langle \psi^{(-1)}_{\nu_B} |\hat a^{\phantom{\dagger}}_{B,j}|\psi^{(0)}_{0} \rangle_B\;.
\end{align*}

Hence the relevance of eigenvectors of the sub-systems for the
partition $l=+1$ depends on the occupation of the border sites in
those states.  We obtain a similar result for $M^{(l=-1)}_\nu$.  From
these considerations we can derive a criterion for the importance of
the contribution of excited eigenvectors of the sub-systems to the
total ground-state of the entire system.

We observe that the partitions $l=\pm 1 $ contribute to second order
energy or rather first order vector correction. The second order
correction for the eigenvectors depend on the partitions $|l|=2$ and
generally the $n$\textsuperscript{th} order correction requires $|l|=n$. This
conclusion can also be obtained by starting a Lanczos iteration with
the eigenvectors of the unperturbed system. Each Lanczos iteration then
increase the required partition $|l|$ by one.

In the Hubbard model we have to deal with the pair
$l=(l_\uparrow,l_\downarrow)$. Starting from $l=(0,0)$, to which the
ground-state of $\hat H_0$ belongs, each Lanczos iteration modifies one of
the values $l_\sigma$ by $\pm 1$. I.e. the first iteration includes
all partitions with Manhattan distance 1 from the center $l=(0,0)$,
the second iteration adds all partitions with Manhattan distance 2 and
so forth.

\section{Block structure of the Hamiltonian}

\begin{figure}
\centering
  \includegraphics[width=5cm]{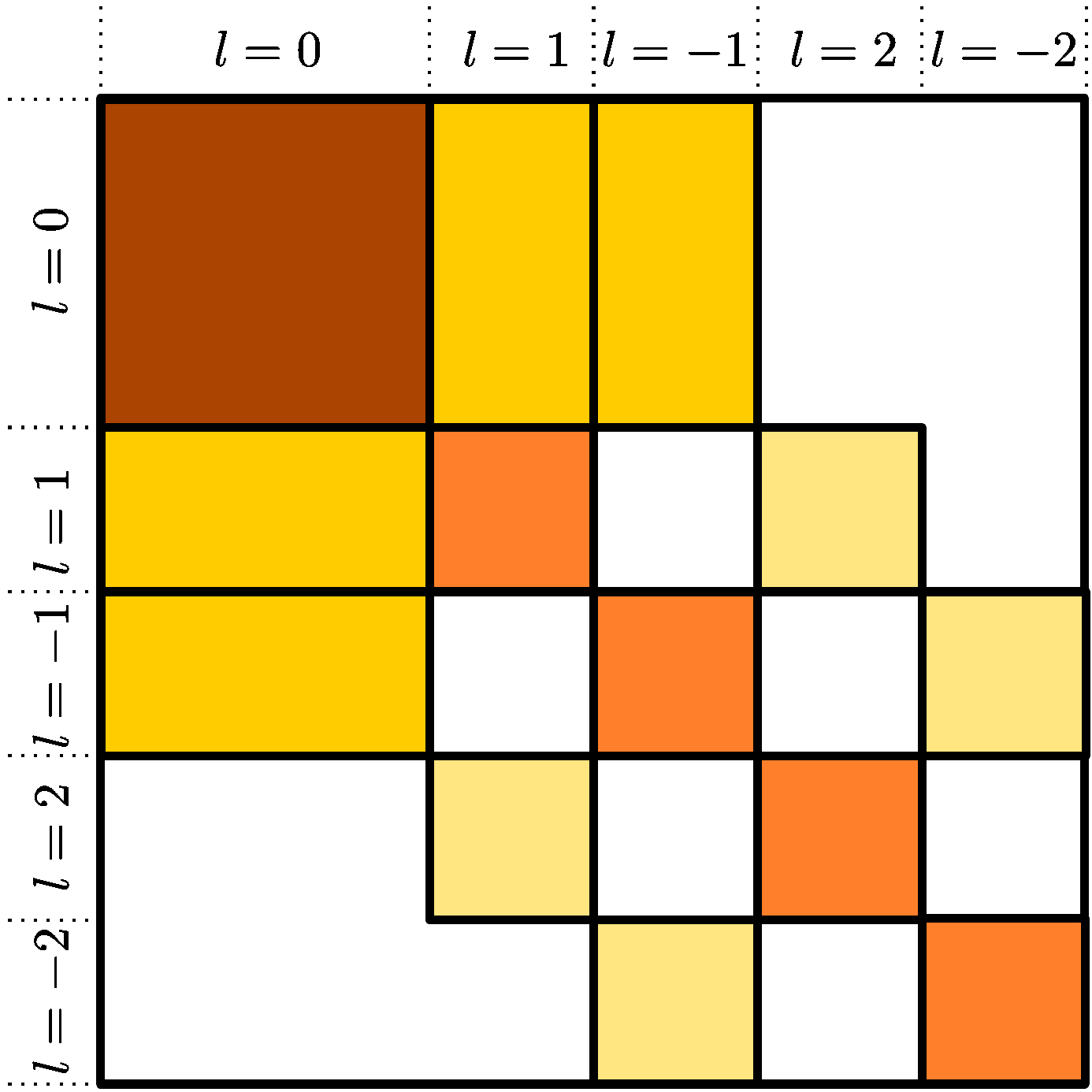} 
\caption{Block-structure of the  Hamiltonian for the  spin-less fermion model.
As explained in the text, $N_A=N_h+l$ and $N_B=N_h-l$.
Only partitions up to  $l=\pm 2$ are depicted. White spaces represent blocks
  with zero matrix elements. \label{TLGSA_matrix_spinless_fermions_3part}}
\end{figure}

The partitioning described in the previous section leads to a natural block
structure of the Hamiltonian, which is depicted for the spin-less fermion model in
fig. \ref{TLGSA_matrix_spinless_fermions_3part}.

The upper-left block corresponds to partition $l=0$ that contains all
many-body basis vectors in occupation number representation
$\ket{\{n\}}$ with equal number of electrons in both sub-systems.

In general the basis vectors in partition $l$ are a tensor product of the contributions of the sub-systems
\begin{align*}
    \ket{n^\lll_\nu} &= \ket{n^\lll_{\nu_A}}_A\otimes \ket{n^\mll_{\nu_B}}_B\;,
\end{align*}
where $\nu=(\nu_A,\nu_B)$ enumerates the basis vectors of the entire
system, while $\nu_\alpha$ enumerates the basis vectors of sub-system
$\alpha$.  The occupation number basis vectors of the two sub-systems
have an outer index $\alpha\in\{A,B\}$ indicating the sub-system they
belong to and they are constructed by the corresponding creation
operators
\begin{align*}
    \ket{n^\lll_{\nu}}_\alpha &= \prod_{i=1}^{L_h} \big(a^\dagger_{\alpha,i}\big)^{n_\nu(\alpha,i)}\;\ket{0}_\alpha\;;\\
    \sum_i n_\nu(\alpha,i) &= N_h\pm l\;;\quad    n_\nu(\alpha,i)\in\{0,1\}\;,
\end{align*}
where $\ket{0}_\alpha$ represents the vacuum vector of sub-system
$\alpha$.  As a consequence of the representation by creation
operators, the two factors of the tensor product do not
commute. Commuting the factors yield an additional sign
$(-1)^{N_A N_B}$. Similarly, there could be an
additional sign when computing matrix elements in the tensor basis.
Due to the tensor structure of the basis the computation of the matrix
elements of $\hat H$ can be simplified significantly.  The contribution of
$\hat H_0$ to the diagonal block of partition $l$ reads

\begin{align*}
  \big(H_0\big)^{(l,l)}_{\nu',\nu} &= \braket{n^\lll_{\nu'_A}| \hat H_A| n^\lll_{\nu_A}}_A\delta_{\nu'_B,\nu_B}\\
  &+\delta_{\nu'_A,\nu_A} \braket{n^\mll_{\nu'_B}| \hat H_B|
    n^\mll_{\nu_B}}_B\;.
\end{align*}

As the diagonal blocks contain a fixed number of electrons in each
sub-system there is no contribution stemming from the hopping part of
$\hat H_{AB}$.  However, in the spin-less fermion case the interaction term
$\hat H_{AB}^\text{int}$ contributes to the diagonal block as well
\begin{align*}
    \left(H^\text{int}_{AB}\right)^{(l,l)}_{\nu',\nu} &= V {\sum_{i j}}'\braket{n^\lll_{\nu_A}|\hat n_{A,i}| n^\lll_{\nu_A}}_A
   \braket{n^\mll_{\nu_B}|\hat n_{B,j}| n^\mll_{\nu_B}}_B\;.
\end{align*}

\begin{figure}
\centering
\includegraphics[width=5cm]{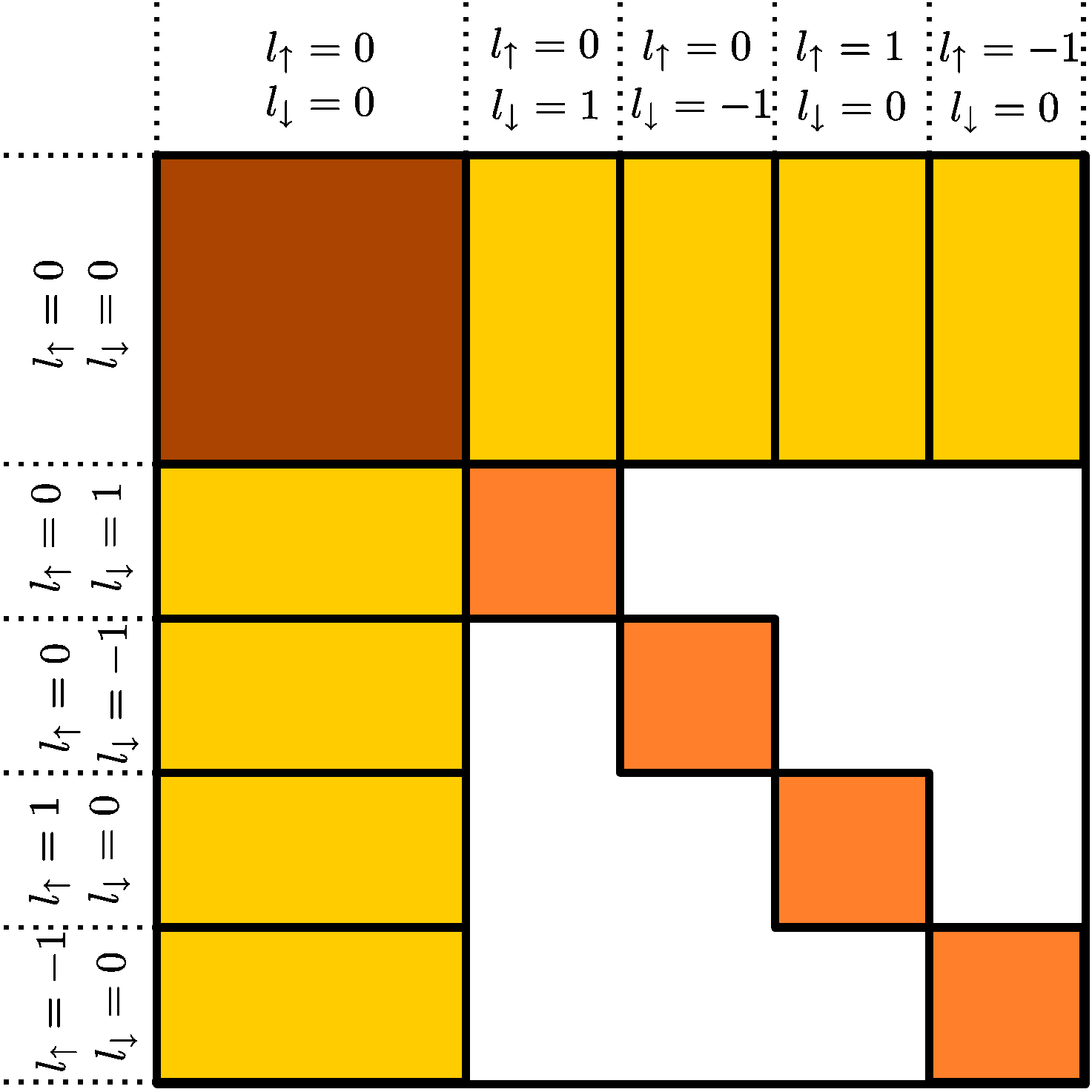} 
\caption{Block-structure of the Hubbard Hamiltonian. The meaning of
  $l_\sigma$ is explained in the text.  Partitions up to Manhattan
  distance 1 are depicted. White spaces represent blocks with zero
  matrix elements. \label{TLGSA_matrix_hubbard_2part}}
\end{figure}

In both models under consideration the off-diagonal blocks are solely
due to the hopping part in $\hat H_{AB}$, which changes the number of
particles in the sub-systems by $\pm 1$. Consequently, only those
blocks possess non-zero entries for which the partition indices differ by
$|l'-l|=1$ in the spin-less fermion case. In the case of the Hubbard
model the condition for non-zero blocks reads
$|l_\uparrow-l'_\uparrow| +|l_\downarrow-l'_\downarrow|= 1$.  I.e. in
the $l_\uparrow,l_\downarrow$ plain only those sites are coupled which
have Manhattan distance 1.  The corresponding spin resolved block
structure is depicted in fig. \ref{TLGSA_matrix_hubbard_2part}.

For the Hubbard model the matrix elements in the off-diagonal blocks with $l'>l$ are
\begin{align*}
  \big(H_{AB}\big)^{(l',l)}_{\nu',\nu}
&= -  t {\sum_{ i,j,\sigma}}'
       \langle n^{(l')}_{\nu'_A} |\hat a^\dagger_{A,i,\sigma} |n^{(l)}_{\nu_A} \rangle_A
       \langle n^{(-l')}_{\nu'_B} |\hat a^{\phantom{\dagger}}_{B,j,\sigma}|n^{(-l)}_{\nu_B} \rangle_B\;.
\end{align*}
Due to the reordering of the creation/annihilation operators
there is an additional sign factor $s=(-1)^{N_h+l}$.
The matrix elements for $l'<l$ follow from the hermiticity $M^{(l,l')}_{\nu,\nu'} = \big(M^{(l',l)}_{\nu',\nu}\big)^\dag$.
The matrix elements for the spin-less fermion case are obtained by restricting the sum over the spins to one spin species, to $\sigma = \uparrow$ say.

All block matrices have the structure
\begin{align}\label{eq:tensor_structure}
    M^{(l',l)}_{\nu',\nu} &= \sum_\kappa \big( M^{(A)}_\kappa \big)^{(l',l)}_{\nu'_A,\nu_A} \big( M^{(B)}_\kappa\big)^{(l',l)}_{\nu'_B,\nu_B}\;,
\end{align}
which will be exploited later on.

\section{Ground-state eigenvalue problem}

In the present approach we first solve the eigenvalue problem of the decoupled sub-systems for the different partitions.
\begin{align*}
    \big( H_0\big)^{(l,l)} V^\lll &= V^\lll D^\lll\;,
\end{align*}
where $V^\lll$ is the unitary matrix of eigenvectors and the diagonal
matrix $D^\lll$ contains the eigenvalues for partition $l$. As before,
in the case of the Hubbard model $l$ stands for
$l=(l_\uparrow,l_\downarrow)$.

We begin with $l=0$ $(l_\uparrow=l_\downarrow=0)$ and include
gradually partitions of increasing Manhattan distance. We always
include all partitions to a a given Manhattan distance, like
$(l_\uparrow,l_\downarrow) \in\{ (1,0), (-1,0), (0,1), (0,-1)\}$. From the
set of eigenvectors $\{\ket{\psi^\lll_\nu}\}$ of these partitions we
keep a certain number, which we call {\it cropping number}.  Next we
form the eigenvector of the entire system by a linear combination of
those selected eigenvectors of the partitions included so far, i.e.
\begin{align}\label{eq:psi_1}
    \ket{\Psi_0^{(l^*)}} &= \sum_{l=0}^{l^*} {\sum_{\nu}} C^{(l^*)}_{l,\nu} \ket{\psi^{(l)}_\nu}\;.
\end{align}
If all partitions and all corresponding eigenvectors are included the result will be exact.
Below, we will give a detailed study of the convergence as far as the upper partition number $l^*$ and the cropping numbers are concerned.

Given  $l^*$ and the set of retained eigenvectors $\{ \ket{\psi^{(l)}_\nu}\}$ for $l\le l^*$, the coefficient vectors $C^{(l^*)}_{l,\nu}$
is given  by the eigenvectors of the matrix
\begin{align}\label{eq:H_tilde}
\tilde H^{(l',l)}_{\nu',\nu} &= \braket{\psi^{(l')}_{\nu'}| \hat H|\psi^{(l)}_\nu}\;,
\end{align}
spanned by the restricted set of unperturbed eigenvectors $\{ \ket{\psi^{(l)}_\nu}\}$ for $l\le l^*$.
The approximated matrix $ \tilde H$
has the same block structure in the partition numbers as the original matrix.
Moreover, the matrix $\tilde H^{(l',l)}$  of block $l',l$  is given by
\begin{align*}
    \tilde H^{(l',l)} &= \big(\tilde V^\llp\big)^\dagger H^{(l',l)} \tilde V^\lll\;,
\end{align*}
where the matrix $\tilde V^\lll$ contains column-wise those the orthonormal eigenvectors of the diagonal block $H^{(0)}_{l,l}$, which are used in the expansion (\ref{eq:psi_1}).

The contribution of $H_0$ to the diagonal block $\tilde H^{(l,l)}$ is diagonal, containing the retained eigenvalues
$D^\lll$.
For the computation of the block matrices resulting from $\hat H_{AB}$ one can exploit the tensor structure outlined in eq.(\ref{eq:tensor_structure})
along with the tensor structure of $\tilde V^\lll$
\begin{align*}
    \tilde V^\lll_{\nu',\nu} &= \tilde V^{(A,l)}_{\nu'_A,\nu_A} \tilde V^{(B,-l)}_{\nu'_B,\nu_B}\;,
\end{align*}
where the columns of $\tilde V^{\alpha,l}$ contain the eigenvectors of $\hat H_\alpha$ retained by the cropping process. So we see that all operations
are restricted to vectors and matrices of the size given by the sub-systems.

\begin{figure}
\centering
\includegraphics[height=6cm]{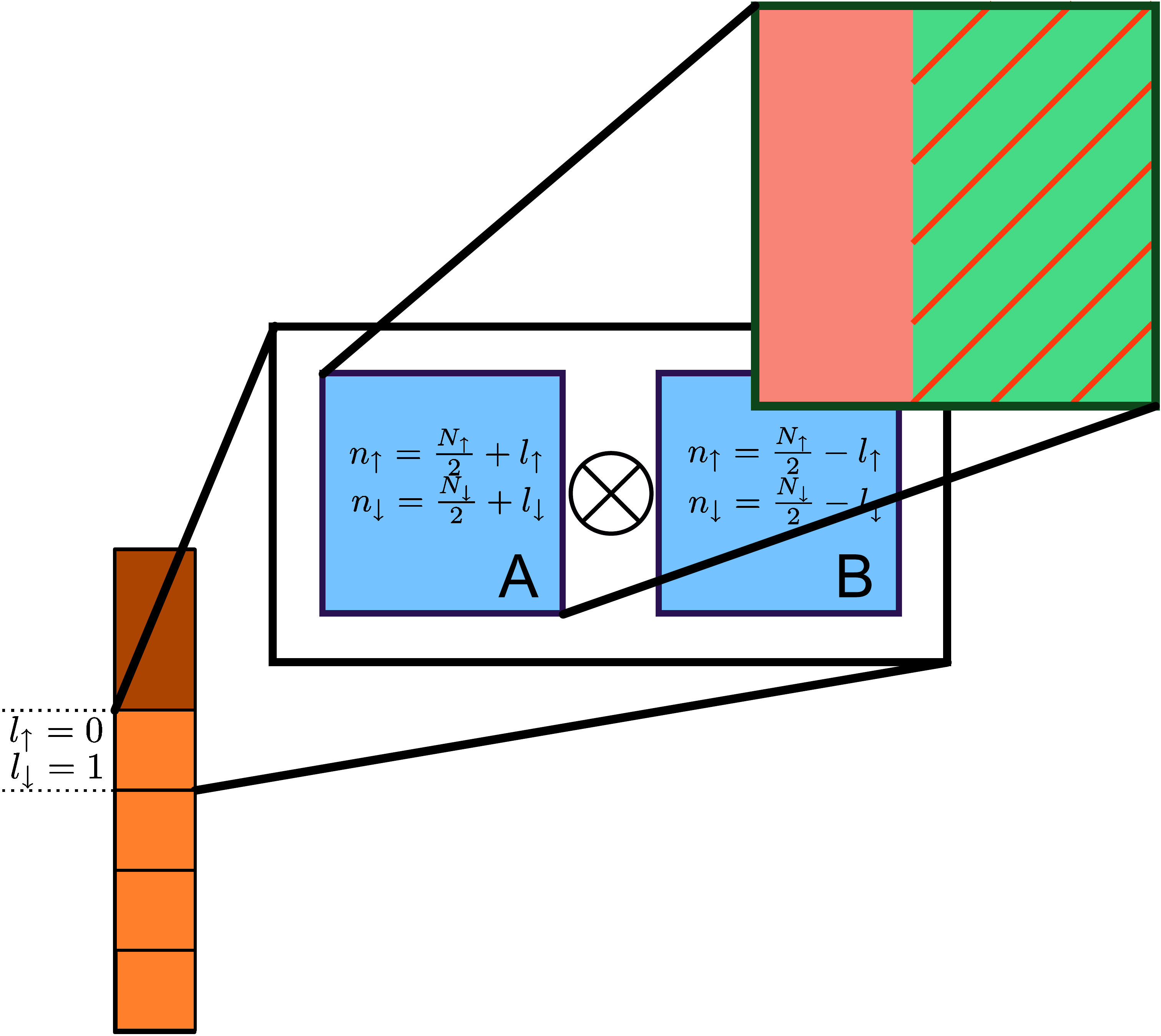}
\caption{Illustration of the construction of the TSGSA basis $\tilde V$. Each partition
  corresponds to a fixed number of spin-up and spin-down particles in both sub-systems, which eigenvectors are combined tensorially (blue) to $\tilde V^{(l)}$. For each sub-system A and B the
  vectors are obtained by solving the corresponding eigenvalue problem.
  The eigenvectors are truncated e.g. by keeping only the vectors on the lower end of the spectrum (red).
  \label{fig:TLGSA_matrix_hubbard_hierachy}}
\end{figure}

Fig. \ref{fig:TLGSA_matrix_hubbard_hierachy} illustrates the
construction of the overall unitary matrix $\tilde V$.

The overall unitary transformation from the orthonormal occupation number basis to the orthonormal basis of the eigenvectors of $H_0$ has a 
block structure, corresponding to the partitions $(l_\uparrow,l_\downarrow)$. The partitions are enumerated  with increasing Manhattan distance and within a shell of fixed distance $d$ counter clockwise beginning with $(l_\uparrow=d,l_\downarrow =0)$. 
\begin{align*}
    V &=
    \begin{pmatrix}
    V^{(1)} & 0 & 0 &\hdots\\
    0& V^{(2)} & 0 & \hdots\\
    0 & 0 & V^{(3)} &\hdots \\
    \vdots&\vdots&&\ddots
    \end{pmatrix}\;,
\end{align*}
where the columns are the eigenvectors of $H_0$ in the various
partitions. Zeros are the corresponding zero matrices.  If we only
retain the restricted set of eigenvalues we obtain a $N_b\times m$
matrix, where $m$ is the number of retained eigenvectors
\begin{align*}
    \tilde V &=
    \begin{pmatrix}
    \tilde V^{(1)} & 0 & 0 &\hdots\\
    0& \tilde V^{(2)} & 0 &\hdots\\
    0 & 0 & \tilde V^{(3)} &\hdots \\
    \vdots&\vdots&&\ddots
    \end{pmatrix}\;,
\end{align*}
the zero matrices are adjusted appropriately. $\tilde V$ is no longer
unitary, but $\tilde V^\dagger \tilde V=I$ still holds for $I$ being
the $m\times m$ identity matrix.  The matrix $P:=\tilde V \tilde
V^\dagger$ is the projection matrix into the space spanned by the
retained eigenvectors. The above procedure corresponds to the
eigenvalue problem of the projected Hamiltonian matrix $\tilde H = P H
P$.

\section{Numerical results}

In order to determine how well the ground-state of a Hubbard model is
approximated by the present approach a series of numerical simulations
was performed and compared to exact values achieved by standard
algorithms like the Lanczos method.

\begin{table}
  \centering
  \begin{tabular}{cccc}
    \hline
     Croppings & $E$   &$|E-E_\text{exact}|$ & rel. error\\
\hline
80/60&-3.114643&0.1177 & 0.0361\\
\hline
80/60/30&-3.221613&0.0108& 0.0033\\
80/60/40&-3.222737&0.0096& 0.0030\\
80/60/50&-3.224861&0.0075& 0.0023\\
\hline
80/60/50/20&-3.226303&0.0060&0.0019 \\
\hline
  \end{tabular}
  \caption{Ground-state energies of a Hubbard system ($L=12, N_\spinup=N_\spindown=6, pbc., U=10$) as estimated by TSGSA compared to Exact Diagonalization (Lanczos). The exact solutions is $E_0=-3.232383$ and the basis size of the full Hamiltonian is 853,776. The first column shows the croppings for the individual partitions used (e.g. for the first line $80$ eigenmodes were used for the primary partition $l=0$; $60$ for the partition $l=\pm 1$, and so on).}
  \label{tab:TLGSA_hubbard_more_partitions_L12}
\end{table}

As can be seen in tab. \ref{tab:TLGSA_hubbard_more_partitions_L12} the
exact eigenvalues are approximated well by a comparably small
vector space. Furthermore, the approximations can be enhanced
by using a limited amount of additional eigenmodes of the sub-systems.

\begin{table}
  \centering
  \begin{tabular}{rrcccc}
    \hline
$L$ & $N_\sigma$ & Croppings &  TSGSA & DMRG & rel. error\\
\hline
4  & 2 &4/2 & -0.882& -0.911 & 0.032\\
8  & 4 & 36/24 & -1.937& -1.975 & 0.019\\
12 & 6 & 50/25& -2.949& -3.041& 0.030\\
16 & 8 &50/25 &  -3.961& -4.109 & 0.036\\
20 & 10 &50/25 &  -5.021& -5.178 & 0.030 \\
24 & 12 &50/25 &  -6.090& -6.245 & 0.025\\
\hline
  \end{tabular}
  \caption{TSGSA for multiple Hubbard system sizes (obc.), compared with DMRG.}
  \label{tab:TLGSA_dmrg}
\end{table}

A study for different system sizes can be seen in
tab. \ref{tab:TLGSA_dmrg}. Here, the DMRG implementation by Reinhard
Noak was used for comparison. Note, that the relative error is stable
to decreasing with system size for constant number of eigenmodes taken
into account. This is due to the fact that with increasing system
sizes the primary partition becomes more important and hoppings over
the sub-system boundary have less weight.

The largest Hubbard-type eigensystem calculation by exact
diagonalization known to the author was preformed at the Earth
Simulator \cite{Yamada_EarthSimulator}. The Hamiltonian used ($L=22$,
$N_\spinup=9$, $N_\spindown=8$) had $1.59 \cdot 10^{11}$ unknowns.
Tab. \ref{tab:TLGSA_large} shows the convergence of a system ($L=22$,
$N_\spinup=N_\spindown=11$) for the for physics important case of
half-filling evaluated by the presented approach. The corresponding
full basis size exceeds the mentioned world-record size by a factor of
3 ($5.0 \cdot 10^{11}$). The calculations were performed using a parallelized
implementation on a 8 QuadCore-Opteron CPU cluster at Graz University
of Technology.

\begin{table}
  \centering
  \begin{tabular}{ccrcc}
    \hline
 Croppings & $N_\text{cores}$ & $t$ [s] & TSGSA & rel. error\\
\hline
25/25/25 & 5&2223&-5.5101& 0.066\\
50/50/50 & 10& 5123&-5.5305& 0.061\\
75/75/75 & 10& 15916&-5.5322& 0.061\\
100/100/100 & 20& 23369& -5.8169&0.013\\
\hline
  \end{tabular}
  \caption{Convergence of the method for a Hubbard-type system ($L=22$, $N_\spinup=N_\spindown=11$, pbc., $U=10$), compared with DMRG (ground-state energy: $-5.8907$). $N_\text{cores}$ indicates the number of computation cores used, $t$ gives the total run-time in seconds.}
  \label{tab:TLGSA_large}
\end{table}

\section{Implementation}

A key to the success of the algorithm is to find a scheme for
calculation of more than just a few lowest eigenvectors, say in the
order of 100. For this purpose an Implicitly Restarted Lanczos
algorithm was used. Note, that the final, transformed Hamiltonian does
not have to be calculated explicitly but a method for its application
on a vector can be derived using the sub-system eigenvalues and
eigenvectors.

Before combining the individual partitions to the effective
Hamiltonian the calculations of the individual sub-systems can be done
completely independent. So a parallelization of the calculation scheme
could be done rather easily by dividing the different sub-system
occupation configurations to different computation cores and solve the
according eigensystems without need for communication. For larger
systems this may not be practical, as the imbalance of numerical
complexities among the different sub-systems makes a straight forward
parallelization more inefficient. For this cases a parallelized
version of the Implicitly Restarted Lanczos using algorithm was
used. 

The algorithm was implemented in the C++ programming language, as a
parallelization framework OpenMPI was used.

\section{Conclusions and Outlook}

The presented approach may lead to a new way of performing
calculations in strongly correlated material sciences. The results are
promising compared to exact diagonalization. Although other
sophisticated methods like DMRG, VCPT, or QMC exist, the TSGSA has the
advantage of producing an explicit matrix representation. Furthermore
it is not limited to one dimensional systems. It can be easily extended
to more dimensions. In this case the off-equilibrium partitions may
become more important due to larger interfaces between the
sub-systems.

A further possibility for developing the algorithm are an intelligent
way of selecting eigenmodes of the sub-system. It can be shown that
many of them do not contribute to the full system ground-state.

\bibliographystyle{model1-num-names}
\bibliography{refs.bib}

\end{document}